\documentclass[useAMS,usenatbib,usegraphicx]{mn2e}
\usepackage{times}

\def\Msun{\rm M_{\rm\odot}}
\def\Rsun{\rm R_{\rm\odot}}
\title[HD 98800]{HD 98800: A most unusual debris disc}
\author[P. E. Verrier and N. W. Evans] 
{P. E. ~Verrier$^1$\thanks{E-mail: pverrier@ast.cam.ac.uk (PEV); nwe@ast.cam.ac.uk (NWE)} 
and N. W. ~Evans$^1$\footnotemark[1]\\
$^1$Institute of Astronomy, University of Cambridge, Madingley Road, Cambridge, CB3 0HA, United Kingdom}
\begin{document}

\date{8 July 2008}

\pagerange{\pageref{firstpage}--\pageref{lastpage}} \pubyear{0000}

\maketitle

\label{firstpage}


\begin{abstract}
  The dynamics of planetesimals in the circumbinary debris disc of the
  quadruple star system HD 98800 are investigated. Evolving a
  spherical shell of test particles from a million years ago to the
  present day indicates that both coplanar and retrograde warped discs
  could exist, as well as a high inclination halo of
  material. Significant gaps are seen in the discs, as well as
  unexpected regions of stability due to the retrograde nature of the
  stellar orbits. Despite a viewing angle almost perpendicular to the
  direction of the warp of the planetesimal disc it is still intersected 
  by the line of sight for eccentricities of the outer orbit of 0.5 or less.
\end{abstract}




\section{Introduction}
\label{sec:intro}

HD 98800 is an interesting and unusual system of four 10 Myr old post
T-Tauri K stars -- two spectroscopic binaries A and B in orbit about
one another -- located in the TW Hydrae association (\citealt{T95},
\citealt{Ka97}). It has a large infrared excess attributed to a
circumbinary disc around the B pair \citep{K00, P01,
  Fu07}. Substantial extinction towards this pair is evidence that
they are observed through some of this material (\citealt{S98},
\citealt{T99}). A photometric variability is also seen, but with no
definite period \citep{S98}. Both the apparent absence of CO molecular
gas in the disc \citep{De05} and infrared spectrum modelling indicate
that this is a T-Tauri transition disc that is just reaching the
debris disc stage, with a collisional cascade having been recently
initiated \citep{Fu07, W07}. The orbits of stars are all highly
eccentric and inclined, creating a dynamical environment unlike almost
all other known debris discs \citep{T99,P01}.

The dust disc here is generally agreed to be an annulus around the B
binary, but the exact structure varies from model to
model. \citet{K00} estimate a coplanar narrow ring outwards of about
$5.0\pm 2.5$ au from the two stars, themselves separated by
approximately 1 au. \citet{P01} however determine a 1 au high coplanar
ring now from 2 and 5 au. \citet{B05} argue that the line of sight
extinction means that the disc is not coplanar with the sub-binary
unless it is very flared. \citet{Fu07} suggest an inner optically thin
ring at 2 au and an outer puffed up optically thick wall 0.75 au high
at 5.9 au, with a gap between the two. Recently, \citet{Ak07} showed
that a single continuous physically and optically thick coplanar disc
between 3 and 10 au can reproduce the observed spectral energy
distribution. To explain the extinction, they used dynamical models of
test particles to show that the inclined stellar orbits could create a
warp in the dust disc, the outer layers of which could then just
intercept the line of sight.

The unusually large infrared excess has been argued by \citet{LBA00}
to be caused by one of four reasons. The first is that a dust avalance
is currently in progress. Radiation pressure acts to push dust grains
outwards. When the disc is very dusty, these can impact other dust
grains, creating more particles that are themselves pushed outwards,
colliding and creating yet more dust, and so on. The second
possibility is there is undetected gas maintaining the dust population
against radiation pressure. The third and fourth reasons are
applicable when the wide orbit has an eccentricity of near unity, much
higher than currently believed. In this case, it is possible that
either an encounter with the A pair has heated the outer edges of the
disc, resulting in abnormally high infrared radiation, or disrupted a
Kupier Belt-like stucture, resulting in collisions and releasing large
amounts of dust. As the eccentricity is now known to be much lower,
these two possibilities are less likely, although it should be noted
that the wide orbit is currently very near to periastron.

The first two possibilities present difficulties in modelling the
system. If the system is undergoing a dust avalance or is gas-rich,
then radiation pressure, collisions and gas dynamics must be included.
For example, \citet{W07} has calculated the dust collision timescale
to very short (0.36 years), so collisional effects are important in
modelling the dust evolution.  However, in debris discs, dust is
generated from an underlying planetesimal population, usually through
a collisional cascade. In this case, the dust population follows the
planetesimal population, which is less complex to model. If the disc
has large amounts of gas or a dust avalance is occuring, this may, of
course, no longer be the case. Models of the planetesimal population
remain interesting though, especially in such an unusual stellar
environment. Indeed, the high inclination and eccentricities suggest
that little stability is likely, yet some must exist. Constraining
possible planetesimal locations and then comparing with observations
of the dust location may even be able to distinguish between the
possible scenarios above. They would also serve as an indication of
whether any planetary stability might be possible in this system.

There is one other debris disc known in a similar stellar system to HD
98800. This is GG Tau, which has a circumbinary debris disc of several
hundred au radius in a quadruple star system with the same hierarchy
\citep{Gu99}. However, in this case the stars are about ten times more
distant, the mass ratios much smaller and their orbits relatively
coplanar so more stability would be
expected~\citep{Be05,Be06}. Dynamical modelling has shown that the
circumbinary material forms a sharp ring and a more diffuse disc
component.

HD 98800 remains an unusual environment, and the high inclination and
eccentricity of the stellar orbits will have a significant effect on
the dynamics and structure of the planetesimal population of the
debris disc.  This effect has yet to be studied in detail, and is
investigated here using direct numerical integrations. Section
\ref{sec:method} gives a description of the stellar system and the
simulation parameters. The results are then presented in Section
\ref{sec:results} and conclusions given in Section
\ref{sec:conclusion}.


\section{Method}
\label{sec:method}

\subsection{The Stellar System}

\begin{figure}
\centering
\includegraphics[width=84mm]{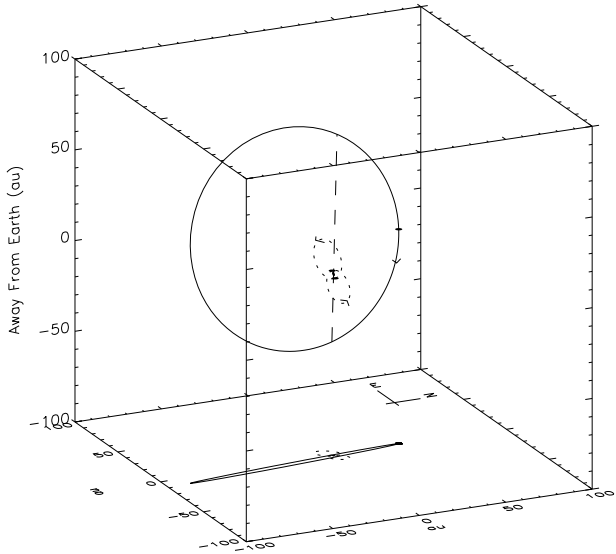}
\caption{
\label{fig:orbit}
The physical orbit of the quadruple star system HD 98800 in three
dimensions. The wide orbit is fit II from \citet{T99} and the B pair
orbit that of \citet{B05}. Units are in au, and the plot is centred on
the barycentre of the B stellar pair. A projection of the orbits onto
the plane of the sky is also plotted along with the relevant
directions. The B pair orbit is overplotted magnified 20 times as
dotted lines. The position of stars at the epoch MJD 52481.34 (mid
2002, inner orbit at periastron) and the direction of the orbits are
given. Finally, the line of intersection of the two orbits is also
plotted as a dashed line.}
\end{figure}

The wide orbit of the sub-binaries A and B is reasonably determined,
as is that of the stars Ba and Bb in the B stellar pair. However the
orbit of the other pair, Aa and Ab, is only partly known as the
smaller star is not resolved. The parameters for all three orbits are
listed in Table \ref{tab:orbit} and shown in Figure \ref{fig:orbit}.

\citet{T99} derives three possible fits to the wide orbit, fixing the
eccentricity in each case as 0.3, 0.5 and 0.6 and assuming a total
system mass of 2.6 $\Msun$. Table \ref{tab:orbit} shows the 0.5
eccentricity case, and Table \ref{tab:ABorbit} shows the others. Apart
from the period (and hence separation), the orbital parameters do not
vary much between the different solutions.

The A pair is a single-lined spectroscopic binary so only a partial
radial velocity orbit has been determined \citep{T95}. \citet{P01} use
evolutionary tracks to estimate the mass of the Aa star as 1.1$\pm$0.1
$\Msun$. The orbital inclination is not known, but the mass and
separation of Ab can be found as a function of this parameter, as
shown in Table \ref{tab:Aab}. A wide range of inclinations are still
possible even given the constraint that the star is small and
unobservable. It is likely that this small star is unimportant to the
dynamics of the circumbinary disc, and so the sub-binary can be
modelled as a single object. However, it might alter the dynamics of
the stellar orbits, and this is investigated in the next sub-section.

\begin{table}
\caption{
\label{tab:orbit}
The orbital and physical parameters for the stellar system HD 98800. The wide orbit of AB is fit II from \citet{T99}, with the semimajor axis calculated assuming a total system mass of 2.6 $\Msun$ and the MJD taken from the middle of the year 2025 given as the time of periastron of the AB pair. The Bab pair orbit is taken from the joint-fit given by \citet{B05}. The orbit of the A sub-binary is from \citet{T95}. Note that the reference plane is that perpendicular to the line of sight and the longitudes of the ascending nodes are measured from North through East.
}
\centerline{
\scriptsize
\begin{tabular}{l|cc@{}c@{}c@{}c@{}c@{}c@{}cc@{}c@{}c@{}c@{}c@{}c@{}c}
Orbital               & Wide   & \multicolumn{7}{c}{A sub-binary}                 & \multicolumn{7}{c}{B sub-binary}                 \\
Parameter             & A      & \multicolumn{3}{c}{Aa} & &\multicolumn{3}{c}{Ab} & \multicolumn{3}{c}{Ba} & &\multicolumn{3}{c}{Bb} \\
\hline
Mass ($\Msun$)        &  --    & 1.1 &$\pm$& 0.1        & &     &     &           & 0.699 &$\pm$& 0.064    & & 0.582 &$\pm$& 0.051   \\
Radius ($\Rsun$)      &  --    &     &     &            & &     &     &           & 1.09  &$\pm$& 0.14     & & 0.85  &$\pm$& 0.11    \\
$a$ (au)              & 67.6   &     &     &            & &     &     &           & 0.447 &$\pm$& 0.013    & & 0.536 &$\pm$& 0.013   \\
Period                & 345 yr & \multicolumn{3}{r@{}}{262.15}  &$\pm$&\multicolumn{3}{@{}l}{0.51 d}  
													& \multicolumn{3}{r@{}}{314.327}  &$\pm$& \multicolumn{3}{@{}l}{0.028 d}	 \\
$e$                   & 0.5    & \multicolumn{3}{r@{}}{0.484}   &$\pm$&\multicolumn{3}{@{}l}{0.020} 
													& \multicolumn{3}{r@{}}{0.7849}   &$\pm$& \multicolumn{3}{@{}l}{0.0053} 	 \\
$i$ (${}^\circ$)      & 88.3   &     &     &            & &     &     &           
													& \multicolumn{3}{r@{}}{66.8}     &$\pm$& \multicolumn{3}{@{}l}{3.2}      	 \\
$\omega$ (${}^\circ$) & 224.6  & \multicolumn{3}{r@{}}{64.4}    &$\pm$&\multicolumn{3}{@{}l}{2.1}    
													& \multicolumn{3}{r@{}}{289.6}    &$\pm$& \multicolumn{3}{@{}l}{1.1}         \\
$\Omega$ (${}^\circ$) & 184.8  &      &     &           & &     &     &                       
													& \multicolumn{3}{r@{}}{337.6}    &$\pm$& \multicolumn{3}{@{}l}{2.4}     	 \\
$\tau$ (MJD)          & 60840  & \multicolumn{3}{r@{}}{8737.1}  &$\pm$&\multicolumn{3}{@{}l}{1.6}  
													& \multicolumn{3}{r@{}}{52481.34} &$\pm$& \multicolumn{3}{@{}l}{0.028}		 \\      
\end{tabular}}
\large
\end{table}

\begin{table}
\caption{
\label{tab:ABorbit}
The three different orbital fits for the wide orbit AB from \citet{T99}. Here $\phi$ is the mutual inclination to the orbit of the B stellar pair.
}
\centerline{
\scriptsize
\begin{tabular}{l|ccc}
Orbital		   		  &	\multicolumn{3}{c}{Orbit} 	\\
Parameter	   		  &	I      &   II   &   III		\\
\hline
$a$ (au)       		  &   61.9 &  67.6  & 78.6      \\
Period (years) 		  &   302  &   345  & 429	    \\
$e$            		  &   0.3  &  0.5   & 0.6	    \\
$i$ (${}^\circ$)      &   87.4 & 88.3   & 88.7      \\
$\omega$ (${}^\circ$) &  210.7 & 224.6  & 224.0     \\
$\Omega$ (${}^\circ$) &  184.8 & 184.8  & 184.8     \\
$\tau$ (MJD)   		  &  59379 & 60840  & 61205     \\
$\phi$ (${}^\circ$)   &  143.0 & 143.7  & 143.9     \\
\end{tabular}}
\large
\end{table}

\begin{table}
\caption{
\label{tab:Aab}
The mass of star Ab and the A binary's semimajor axes as a function of inclination to the plane of the sky.
}
\centerline{
\scriptsize
\begin{tabular}{l|llllllll}
Inclination (${}^\circ$)   & 90 & 70 & 60 & 50 & 40 & 30 & 20 & 10 \\
\hline
$M_{Ab}$ ($\Msun$)  & 0.22 & 0.23 & 0.25 & 0.29 & 0.36 & 0.49 & 0.80 & 2.36 \\
$a$ (au)            & 0.88 & 0.88 & 0.89 & 0.89 & 0.91 & 0.94 & 0.99 & 1.21 \\
\end{tabular}}
\large
\end{table}

\subsection{The Stellar Dynamics}

The mutual inclination for the stellar orbits given in
Table \ref{tab:orbit} can be calculated using the formula
\begin{equation}
\cos \phi = \cos i_1 \cos i_2 + \sin i_1 \sin i_2 \cos(\Omega_2-\Omega_1)
\end{equation}
where $\Omega_j$ are the longitudes of ascending node, $i_j$ are the
inclinations relative to a given reference plane and $\phi$ is the
mutual inclination, the angle between the angular momentum vectors of
the two orbits (see e.g. \citealt{S53}). It is found to be in this
case $143.7^\circ$, and is similar for the other two orbits (see Table
\ref{tab:ABorbit}). Notably, it is retrograde.

\begin{figure}
\centering
\includegraphics[width=3.15in]{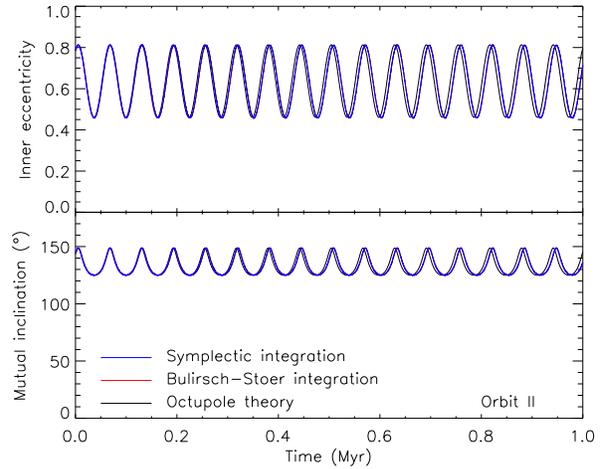}
\caption{
\label{fig:3stars}
The three body stellar orbital evolution for orbit II over 1 Myr from
numerical and theoretical modelling. The blue line shows the results
from the symplectic integrator, identical to those from the
Bulirsch-Stoer shown in red. The octopole theory results are shown in
black, and differ only slightly in phase. The semimajor axes of both
orbits and eccentricity of the outer orbit are constant and not
shown.}
\end{figure}

As mentioned, it is likely for an investigation of the dynamics of the
circumbinary disc that the Aab system can be reasonably approximated
by a single star of mass 1.3 $\Msun$, consistent with the minimum mass
of Ab and the total system mass assumed by \citet{T99}. In this case,
the system becomes a hierarchical triple and can be numerically
studied using the {\sc{Moirai}} code \citep{VE07}. A million year
simulation of orbit II in this case is shown in Figure
\ref{fig:3stars}. Energy is conserved to a relative error of $10^{-7}$
and the integration was checked with a standard Bulirsch-Stoer
integrator \citep{P89} and found to be in agreement. The semimajor
axes of the stars and the wide orbit's eccentricity are not shown, as
they remain constant through out the simulation. Interestingly, the
system is currently near its maximum eccentricity and mutual
inclination. For all three possible outer orbits, the eccentricity and
mutual inclination vary smoothly in a secular manner, with maximum
angular separation between the orbital planes occurring for minimum
eccentricity of the inner orbit. The amplitude of the variation is
very similar in all cases and the periods are 65, 60 and 75 Kyr for
orbits I, II and III respectively. In fact the system remains in this
stable configuration for at least a Gyr, as well as to at least 10 Myr
ago.

The orbital behaviour is well described by the octupole secular theory
of \citet{FKR00}, as overplotted in Figure \ref{fig:3stars}. This
theory uses a third order expansion of the systems Hamiltonian in the
ratio of the semimajor axes of the stellar orbits to obtain a set of
coupled first order differential equations, their equations (29) to
(32), for the time variation of the eccentricities and arguments of
periastron of the inner and outer orbits that can be numerically
solved. The inclination and nodes of the system are then derivable
from these elements, and the semimajor axes remain constant. From the
figure, it can be seen that the theory is in excellent agreement with
the results from the full equations of motion, with only a small phase
drift after a Myr. In fact, because the two orbits are fairly
separated a quadrupole level theory is sufficient to describe the
system. In this approximation the outer eccentricity is a constant,
as seen here. Thus, in the three body approximation the stars follow a
stable secular evolution and can be accurately integrated by the
{\sc{Moirai}} code.

The full four body system can now be considered.  The mass of Ab and
the semimajor axis of the orbit can only be resolved by assuming the
inclination to the line of sight (see Table \ref{tab:Aab}). The
longitude of ascending node of the orbit remains an unknown, but is
needed to constrain the mutual inclination of the orbital
planes. Therefore, to investigate possible dynamics, a set of
simulations were run for inclinations in the range $90^\circ$ to
$30^\circ$ (shown in Table \ref{tab:Aab}) with values of ascending
node ranging from $0^\circ$ to $315^\circ$ in $45^\circ$ steps and
using wide orbit II. To do this, the fourth star was approximated as a
planet around its primary, Aa, in the {\sc{Moirai}} code.

\begin{figure}
\centering
\includegraphics[width=3.15in]{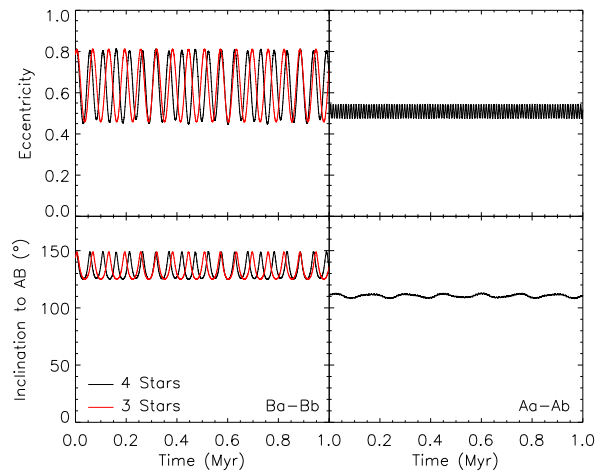}
\caption{
\label{fig:muti}
The evolution of the stellar orbits for the four body case with
$i_{Ab}=30^\circ$ (relative to the plane of the sky) and
$\Omega_{Ab}=45^\circ$ compared to the three body approximation. The
left hand panels show the orbit of the inner binary Ba-Bb, red
indicating the three body approximation and black the four body
case. The right panels show the evolution of the orbit of the Aa-Ab
binary. The eccentricity of the wide binary A-B once again remains
effectively constant, as do the semimajor axes of all orbits.}
\end{figure}

In all cases, there is little difference to the three body results. An
example is shown in Figure \ref{fig:muti}. The only change is to the
secular period of orbit Ba-Bb together with a slight modulation of
their minimum eccentricity, and in some cases even this does not
occur. A slight change in the period of the secular variations is
unlikely to alter the overall structure of the planetesimal disc. Indeed,
\citet{Be06} also find that modelling the distant sub-binary in GG Tau
as a single object has little effect on the disc structure
there. Hence, the three body approximation is a reasonable assumption
and will be used for the purposes of this investigation.

\subsection{Other Physics}

As the disc mass is low (0.002$\rm{M}_\oplus$, \citealt{P01}), it is
reasonable to model the planetesimals as non-interacting test
particles.  However, there is one interaction process that may be
important in the modelling of the planetesimals, namely
collisions. Following \citet{W07}, it is possible to esimate the
collision timescale from the observed infrared excess.  Although this
assumes a collisional cascade is underway, it should provide a lower
limit to the timescale if dust is in fact maintained by gas or
generated in an avalance.  The collision timescale is given by their
equation (13) as
\begin{equation}
t_c = \left( \frac{3.8\rho r^{2.5} dr D_c}{M_{\star}^{0.5}M_{tot}}\right)
       \left( \frac{(12q-20)(1+1.25(e/I)^2)^{-0.5})}{(18-9q)G(q,X_c)} \right)
\end{equation}
where $q$ is assumed to be $11/6$ for a collisional cascade and $G$ is
a function of $q$ and $X_c=D_{cc}/D_c$, $D_c$ being the diameter of
largest planetesimal in the cascade, taken as $2000$ km, and $D_{cc}$
being the smallest planetesimal that has enough energy to destroy
another of size $D_c$. $\rho$ is the planetesimal density and taken as
$2700$ kg m${}^{-3}$, $r$ is the planetesimal belt radius in au, $dr$
is the planetesimal belt width in au, $M_{tot}$ is the total mass of
material in the cascade in $M_{\oplus}$, $e$ is the mean orbital
eccentricity of planetesimals and $I$ is the mean orbital inclination
of planetesimals in radians. The factor $G(q,X_c)$ is given by their
equation (9) as
\begin{eqnarray}
G(q,X_c) &=& (X_c^{5-3q}-1) + \frac{6q-10}{3q-4}(X_c^{4-3q}-1) \nonumber \\
{} & &    + \frac{3q-5}{3q-3}(X_c^{3-3q}-1) 
\end{eqnarray}
and $X_c$ by their equation (11) as
\begin{equation}
X_c = 1.3 \times 10^{-3} \left( Q_D^{\star} r M_{\star}^{-1}f(e,I)^{-2} \right)^{1/3}
\end{equation}
where $f(e,I)$ is the ratio of the relative collision velocity to the
Keplerian velocity and taken as 0.1 and $Q_D^{\star}$ is the dispersal
threshold which is the specific incident energy needed to destroy a
particle, assumed to be 200 J kg${}^{-1}$. In addition, it is also
assumed that $e \approx I$.

The two stars Ba and Bb are approximated as a single object of mass
and luminosity as 1.57 $M_\odot$ and 0.58 $L_{\odot}$ respectively
\citep{P01}. The total mass of the disc can be calculated from
equations (4) to (6) of \citet{W07} as
\begin{equation}
M_{tot} = 5.194 \sigma_{tot} =  14.36 r^2
\end{equation}
and hence the collision timescale as a function of $r$ and $dr$ is
\begin{eqnarray}
t_c &=& 1.014 \times 10^{6} r^{0.5} dr \frac{1}{G(q,X_c)}
\end{eqnarray}
The exact extent of the planetesimal disc is as yet unknown, but the
location of the dust can be used as an approximation.  The minimum
radius for dust to be able to exist is given as 2.2 au by \citet{L05},
but \citet{K00} estimates the disc to be between 5.0 and 18 au. In the
first case of a wide disc around 2 au, the collision timescale is 30
Kyr, and in the second case of a disc from 5 to 18 au, it is 200
Kyr. Preliminary results from test simulations indicated that the disc
must lie further out than 2 au, so the higher timescale is applicable
and gravitational effects will dominate the behaviour of the
planetesimals here. Thus, for the purpose of determining overall disc
structure, no additional physics needs to be included.

\subsection{Method of investigation}

To summarise, the stars are modelled as a three body system with a
disc of massless planetesimals interacting through gravitational
effects with the stars only. The three possible orbital configurations
for the wide binary will all be considered. To model the disc, the
{\sc{Moirai}} code has been shown to be accurate, and the test
particle disc can be implemented as circumbinary particles around the
B stellar pair.

The models discussed in the introduction have all assumed that the
disc is coplanar with the B binary, but there is evidence for inclined
discs around similar multiple stars. For example, the close binary
pair in the T Tauri system were recently determined to have misaligned
circumstellar discs \citep{Sk07}, and polarization surveys have found
a small number of similar cases \citep{Mo06}. Because of this, the
test particle distribution is not initially taken as
coplanar. Instead, they are spaced uniformly in inclination and
longitude relative to the B binary pair. The initial surface density
of the disc is taken as $1/r$ and the grid of test particles runs from
1.75 au to 33.85 au (inner binaries radius to half the outer binaries
radius), as shown in Table \ref{tab:grids}. As the present day
configuration, stability and geometry of the system are of interest to
compare to observations, the simulations are run from 1 Myr ago to
present day (defined as the periastron time of the Ba-Bb pair from
Table \ref{tab:orbit}). The results from these simulations are
presented in the next section.

\begin{table}
\caption{
\label{tab:grids}
   The test particle grid used in the simulations.
}
\centerline{
\scriptsize
\begin{tabular}{l|ccc}
Orbital Parameter    &    Min  &   Max   &  Step Size   \\
\hline
$a$ (au)			 &   1.75  &  33.85  &  0.1    \\
$e$					 &   0.0   &   0.04  &  0.02    \\
$i$ (${}^\circ$)	 &   0     &   180   &  5       \\
$\omega$ (${}^\circ$)&   0     &   240   &  120     \\
$\Omega$ (${}^\circ$)&   0     &   240   &  120     \\ 
$M$ (${}^\circ$)	 &   0     &   240   &  120     \\
\hline
$\rm{N_{tp}}$			 & \multicolumn{3}{c}{965034}  
\end{tabular}}	
\large
\end{table}


\section{Results}
\label{sec:results}

\begin{figure}
\centering
\includegraphics[width=3.15in]{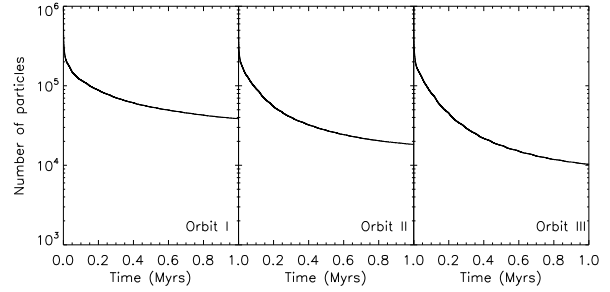}
\caption{
\label{fig:decay}
The test particle decay rates for the three simulations. The half-life
(ignoring the first 0.1 Myr when all initially unstable particles are
removed) is about 0.3 Myr for orbit I, 0.2 Myr for orbit II and 0.1
Myr for orbit III.}
\end{figure}

The stability of planetesimals is indicated by the stability of the test
particles. These are removed from the simulation if they cross
either of the stellar orbits or if they become unbound. The test
particle decay rates are shown in Figure \ref{fig:decay} for the three
different simulations. The half-life is short and by the end of the
simulations the decay rate has levelled out, although some particles
are still being slowly eroded from very unstable locations. The
majority of unstable particles are removed on 10 to 100 Kyr timescales
as they become unbound or cross the outer orbit. If the simulations
are run for an additional million years, there is no further
significant loss of particles so the simulation length is sufficient
to determine the system state, especially given the system age of
about 10 Myr.

\begin{figure*}
\centering
\includegraphics[width=7in]{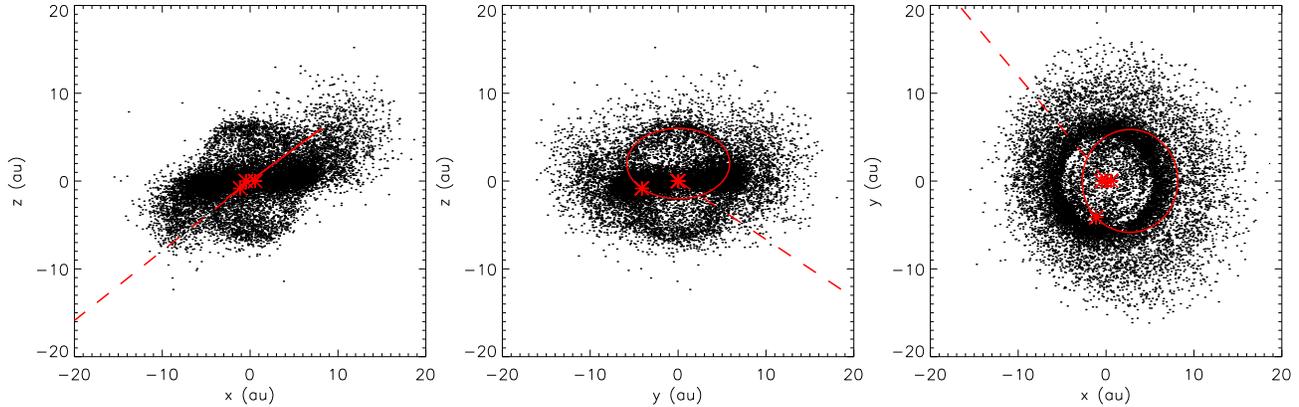}
\caption{
\label{fig:II_faceon}
Edge and face on views of the circumbinary material for the simulation
using orbit II. The inner binary's orbit lies in the $z=0$ plane and
the line of intersection with the outer stellar orbit is along the
$y$-axis. The orbits of the stars and their current positions are
overplotted in red, but the wide orbit is shown at a tenth of its
actual size. The line of sight is also shown as a dashed line.}
\end{figure*}

\begin{figure*}
\centering
\includegraphics[width=7in]{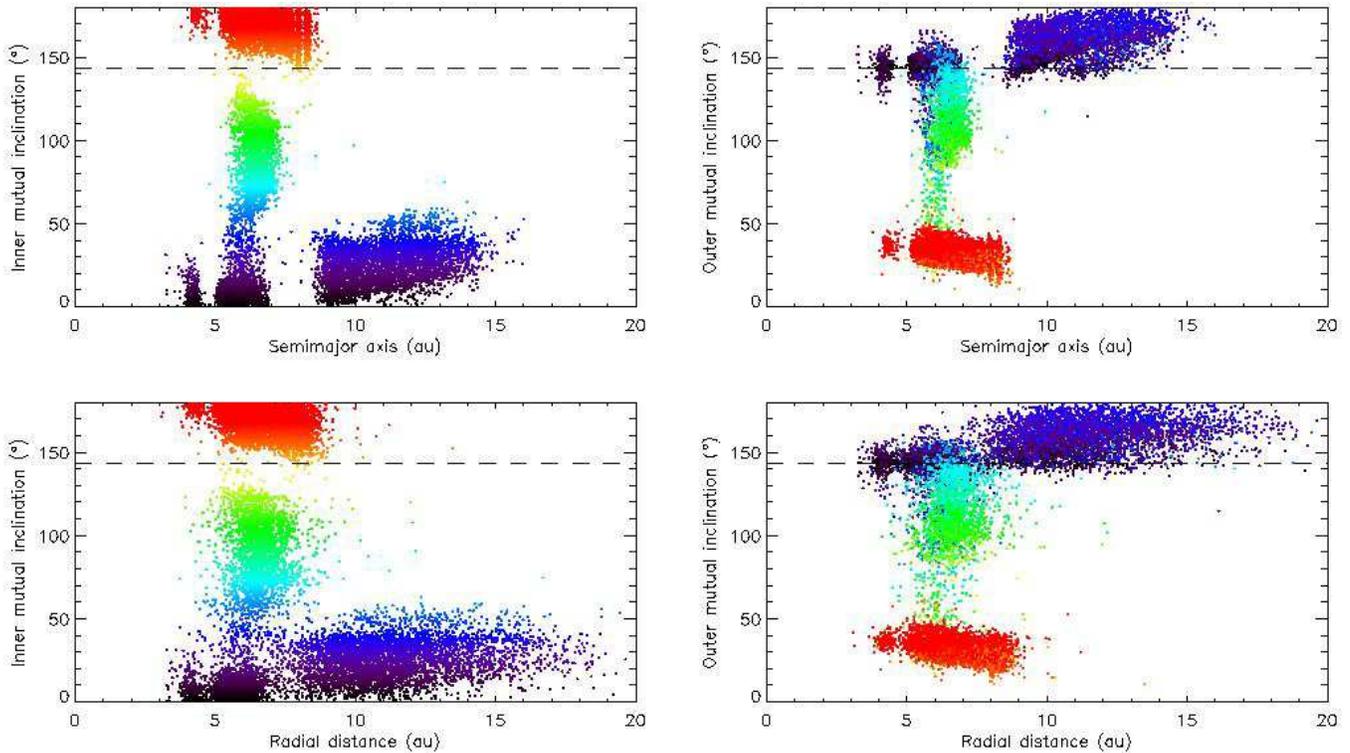}
\caption{
\label{fig:II_graphs}
The final test particle distributions for the simulation using orbit
II. The two panels on the left show the inclination relative to the
inner binary as a function of semimajor axis and radial distance from
the center of the binary, with colour indicating inner mutual
inclination as shown. The two panels on the right show the inclination
relative to the wide orbit, but the colour still indicates inner
inclination. Overplotted as a dashed line is the mutual inclination of
the two stellar orbits.}
\end{figure*}

The final structure at the end of the simulation is similar in all
three orbital cases, and is well illustrated by the case of orbit
II. Shown in Figure \ref{fig:II_faceon} is the spatial distribution of
the circumbinary material. A warped coplanar disc is apparent around
the inner binary, as well as a large amount of higher inclination
material. The disc appears to have a small inner ring separate to the
main bulk of particles. There is also a slight clumping of material
apparent, perpendicular to the inner orbits line of apses. The high
inclination material appears to form another ring or halo
perpendicular to the disc, most obvious in the middle plot of the
figure.

These different structures are more clearly seen in a plot of
inclination as a function of distance from the inner binary, as shown
in Figure \ref{fig:II_graphs}. Here, the particle distribution
relative to the inner and outer orbits is plotted for both radial distance and
semimajor axis. To identify different populations, particles
are colour coded to their inclination relative to the inner binary's
orbit.

Three distinct features are seen in these plots: a prograde coplanar
disc with two gaps; a retrograde disc with one gap and an extended
ring or halo, as seen in the spatial plot, from $50^\circ$ to
$130^\circ$ inner inclination. These populations are well separated in
inclination by two gaps around $50^\circ$ and $140^\circ$, the later
being centred on the inclination of the outer binary's orbit.

\begin{figure*}
\centering
\includegraphics[width=7in]{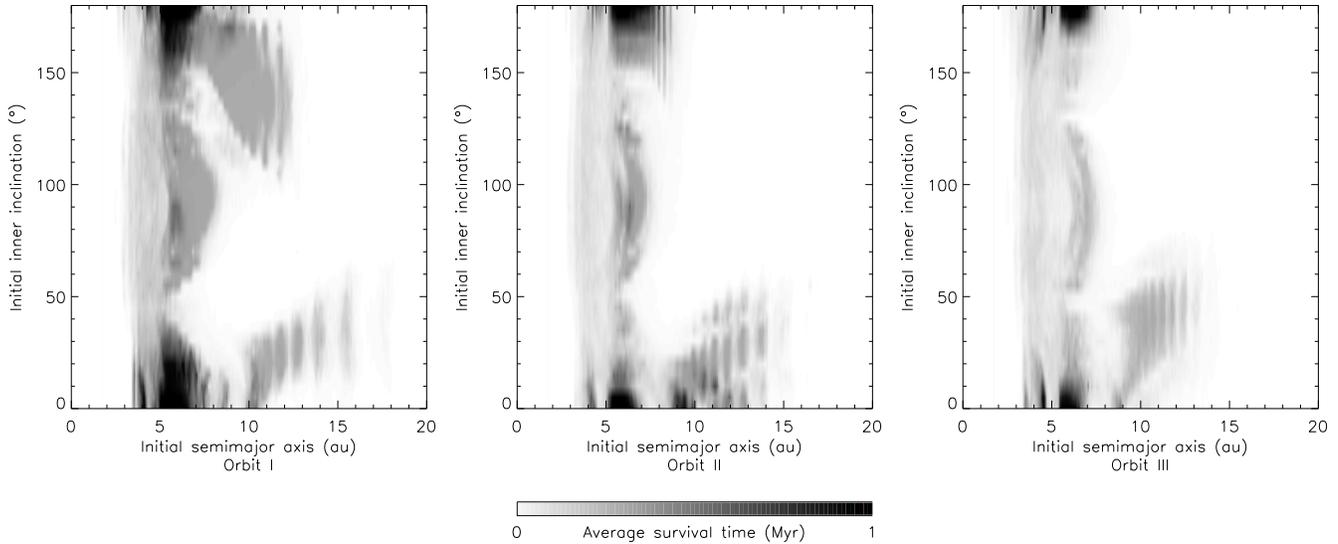}
\caption{
\label{fig:stabmap}
Stability maps comparing the results from the three simulations using orbits I, II and III. The average survival time is shown as a function of initial semimajor axis and inner inclination, with black indicating all particles starting at a given grid point remained at the end of the simulation through to white indicating the location was very quickly unstable.}
\end{figure*}

The coplanar disc extends from about 4 to 15 au, with two gaps between
4.5 and 5 au (seen very clearly in Fig. \ref{fig:II_faceon}) and 7 and
8 au. The gaps are less obvious in the radial distribution due to
particle eccentricity, but are still apparent. The outer regions of
this disc are inclined up to almost $50^\circ$ and provide the warped
material seen in the spatial plot. There is also a series of breaks in
the semimajor axis distribution slightly apparent both here and in the
retrograde disc (these are very clear in Figure~\ref{fig:stabmap}),
presumably due to resonant features as they are much larger than the
grid resolution of 0.1 au. Relative to the outer orbit, the disc
material in the outer warped regions is perturbed up to almost
completely retrograde orbits, explaining its stability.

The retrograde (relative to the inner binary) disc is smaller,
extending from 4 to 9 au, but also has a gap between 4.5 and 5 au. The
outer regions here are perturbed towards the same inclination as the
outer orbit, similar but opposite to the coplanar disc's
structure. This disc is in fact much sharper relative to the outer
star, where it is prograde with an inclination of around $30^\circ$.

The last component is the halo-like structure. This extends only from
5.0 to 8.0 au but covers a large range of inclinations relative to
both stellar orbits. This material is unusual in remaining stable at
very high inclinations, but similar stable high inclination particles
have been seen before by \citet{Be06} in simulations of GG Tau.

\begin{figure}
\centering
\includegraphics[width=3.15in]{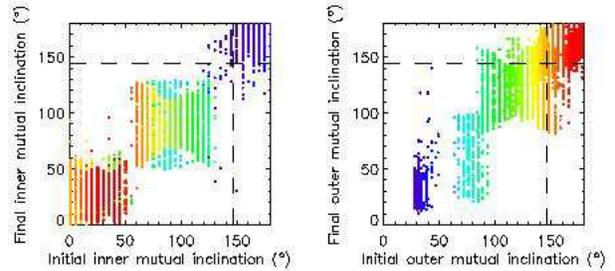}
\caption{
\label{fig:II_initial}
The initial test particle inclinations related to their final
inclinations for the simulation using orbit II. The left panel shows
inclination relative to the inner binary and the right panel relative
to the outer wide orbit. Colour now indicates initial outer
inclination to highlight the small population around $80^\circ$ in the
right hand plot. The dotted vertical line shows the initial mutual
stellar inclination and the horizontal the final.}
\end{figure}

These three separate populations are also very distinct in a plot of
final versus initial inclination, as shown in Figure
\ref{fig:II_initial}. Particles in the prograde disc start with
initial inner inclinations in the range $0^\circ$ to $50^\circ$ and
remain within that range. Those in the halo, starting in the range
$60^\circ$ to $125^\circ$, also remain there although there is less
variation in the middle of this group. Finally, the retrograde disc is
clearly confined by the inclination of the outer stellar orbit, and
particles here start and remain in the range $145^\circ$ to
$180^\circ$. These populations are again clear in the outer
inclination distribution, as can be seen from the colour coding of the
particles. There is in fact a fourth diffuse population visible here,
at around $70^\circ$ to $90^\circ$, which forms a more diffuse ring at
a different angle to that of the main halo.

The initial outer inclination distribution is different for the orbit
I and III simulations, as the stars start in slightly different
places. However, the components evolve to the same final inclination
distribution, showing that the results are robust and do not strongly
depend on parameters such as the initial longitudes. Note also that in
the orbit II case no particles start with initial outer inclinations
less than about $30^\circ$, so it could be possible that another
region of stability exists here. However, a disc of particles started
coplanar with this outer orbit very rapidly became unstable, with no
material surviving to the end of the simulation.

\begin{figure*}
\centering
\includegraphics[width=7in]{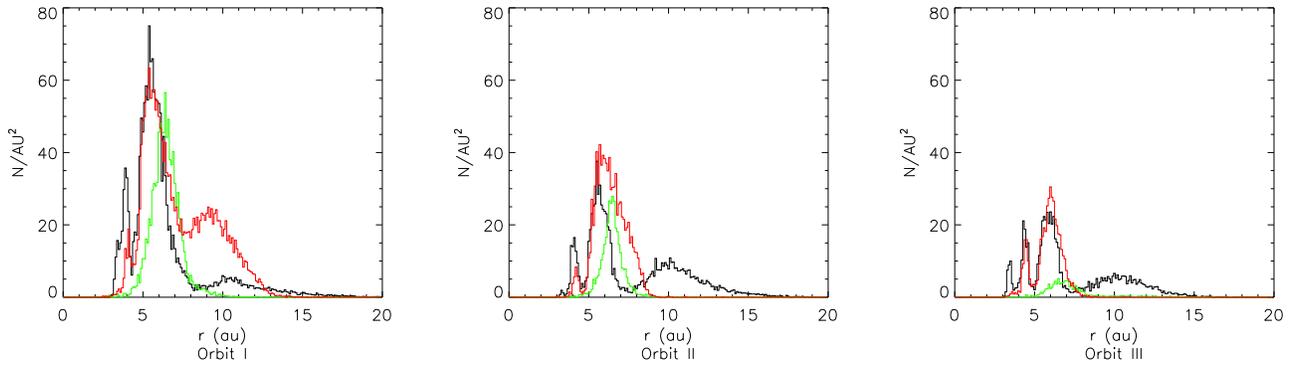}
\caption{
\label{fig:II_profi}
The radial profile of the three inclination components of the
circumbinary material for all three simulations, showing the surface
density as a function of radius. The coplanar disc is plotted in
black, the halo in green and the retrograde disc in red. Note that for
the orbit I case the retrograde disc starts at $105^\circ$ instead of
$145^\circ$ due to the differences in the initial inclinations of the
stars for this orbit. The retrograde disc and halo here also now
overlap in inner inclination, and are separated using their outer
inclination instead.}
\end{figure*}

A radial profile of the three main structures can be plotted by using
the inner inclination range to characterise them. This is shown in
Figure \ref{fig:II_profi} for each simulation. The coplanar disc
(black) is defined as starting between $0^\circ$ to $50^\circ$, the
halo (green) between $55^\circ$ to $140^\circ$ and the retrograde disc
between $145^\circ$ to $180^\circ$. In each case the profiles of each
component are very different, and the gaps in the two disc very
apparent.

The radial profiles here are the result of the evolution of an initial
$1/r$ distribution. A simulation was rerun for a flat initial density
profile in the orbit II case. The resulting disc profiles were
similar, although the inner and outer edges were slightly steeper. The
profile of the halo also remained largely unchanged.

The most notable difference between the three simulations is the
generally greater stability in the lower eccentricity cases and the
existence of an extended retrograde disc in the orbit I case. This
stability is shown in greater detail in Figure \ref{fig:stabmap}, and
compared to the other simulations. As the extended prograde coplanar
disc is retrograde relative to the outer star, this disc is prograde
to it. The extended prograde disc is itself less populated here, but
both features can partly be explained by the initial particle
inclinations. The stellar orbits are different here and particles that
are retrograde relative to the inner star are more coplanar with the
outer star than in the other orbital cases, and those that have
prograde inner inclinations have lower outer inclinations. Thus, fewer
particles start in the prograde extended disc, while more start in the
retrograde case. It should be noted, however, that even if the initial
inclination distribution is similar to the other two simulations, a
retrograde extended disc is still seen. The eccentricity of the
stellar orbits is therefore still likely to be important to these
features, as can be seen by the decreasing radial extent of the
prograde disc from orbit I to III in Figure \ref{fig:stabmap}. Indeed,
as discussed above, a retrograde disc relative to the outer star for
the orbit II case is completely unstable.

The radial extent of the structures seen in the three simulations is
detailed in Table \ref{tab:rlimits} and compared to both previous
estimates of the size of the observed dust disc and empirical
stability limits from numerical studies of general binary and triple
systems. These empirical limits show the inner and outer radius for
coplanar circumbinary stability by modelling the stars as a low
inclination triple system \citep{VE07} and as two coplanar decoupled
binary systems \citep{HW99}. A more general result from \citet{VE07}
is that in cases of high stellar eccentricity and mass ratio the
stable region was more complex than a completely stable ring, with
gaps appearing that were most likely due to overlapping resonances
from the two stellar orbits, as seen here. The extended outer regions
of the prograde and retrograde discs are outside the empirically
predicted stable region. This is most likely an effect due to the high
inclinations and the particles retrograde nature relative to the outer
or inner orbit, as retrograde orbits are generally more stable than
prograde. However, the inner edges are well predicted.

\begin{table}
\caption{
\label{tab:rlimits}
The radial extent of the disc around HD 98800 B. The top four rows show other models of this system, the rest show empirical fits from general binary and triple systems and the extent of the three different inclination components for each possible outer orbit from the simulations here. 
}
\centerline{
\scriptsize
\begin{tabular}{l|l@{\hspace{0.05in}}c@{\hspace{0.05in}}l@{\hspace{0.05in}}c@{\hspace{0.05in}}l@{\hspace{0.05in}}c@{\hspace{0.05in}}r}
Model 		& \multicolumn{4}{l}{Inner edge (au)} 	 & \multicolumn{3}{r}{Outer edge (au)}	\\
\hline
\citet{K00} & \multicolumn{7}{c}{$5.0 \pm 2.5$}												\\
\citet{P01} & 2.0	     &     &               & --  &            &	    & 5.0				\\
\citet{Fu07}& 2.0        & \multicolumn{5}{c}{gap}                      & 5.9				\\
\citet{Ak07}& 3.0 	     &     &               & --  &            & 	& 10.0				\\
\hline
\multicolumn{8}{c}{Orbit I}																	\\
\hline
\citet{HW99}& \multicolumn{3}{l}{4.1}& --  & \multicolumn{3}{r}{10.9}	\\ 
\citet{VE07}& \multicolumn{3}{l}{3.9}& --  & \multicolumn{3}{r}{11.0}	\\
\hline
Disc		& 3.0 -- 4.0 &     &               & gap & 4.5 -- 8.0 & gap & 9.0 -- 20.0		\\
Halo    	& 5.0		 &     &               & --  &            & 	& 8.0				\\
Retrograde  & 3.5 -- 4.0 &     &               & gap & 4.5 -- 7.0 & gap & 7.5 -- 13.0 		\\
\hline
\multicolumn{8}{c}{Orbit II}																\\
\hline
\citet{HW99}& \multicolumn{3}{l}{4.1}& --  & \multicolumn{3}{r}{4.3}	\\ 
\citet{VE07}& \multicolumn{3}{l}{3.9}& --  & \multicolumn{3}{r}{0.6}	\\
\hline
Disc		& 4.0 -- 4.5 &     &               & gap & 5.0 -- 7.0 & gap & 8.0 -- 15.0 		\\
Halo    	& 5.0		 &     &               & --  &            &  	& 8.0				\\
Retrograde  & 4.0 -- 4.5 &     &               & gap & 5.0        & --  & 9.0 				\\
\hline
\multicolumn{8}{c}{Orbit III}																\\
\hline
\citet{HW99}& \multicolumn{3}{l}{4.1}& --  & \multicolumn{3}{r}{7.0}	\\ 
\citet{VE07}& \multicolumn{3}{l}{3.9}& --  & \multicolumn{3}{r}{7.3}	\\
\hline
Disc		& 3.0 -- 3.5 & gap & 4.0 -- 4.5    & gap & 5.0 -- 7.0 & gap & 8.0 -- 15.0		\\
Halo    	& 5.0		 &     &               & --  &            &     & 8.0				\\
Retrograde  & 4.0 -- 4.5 &     &               & gap & 5.0        & --  & 8.0	
\end{tabular}}	
\large
\end{table}

\begin{figure*}
\centering
\includegraphics[width=4in]{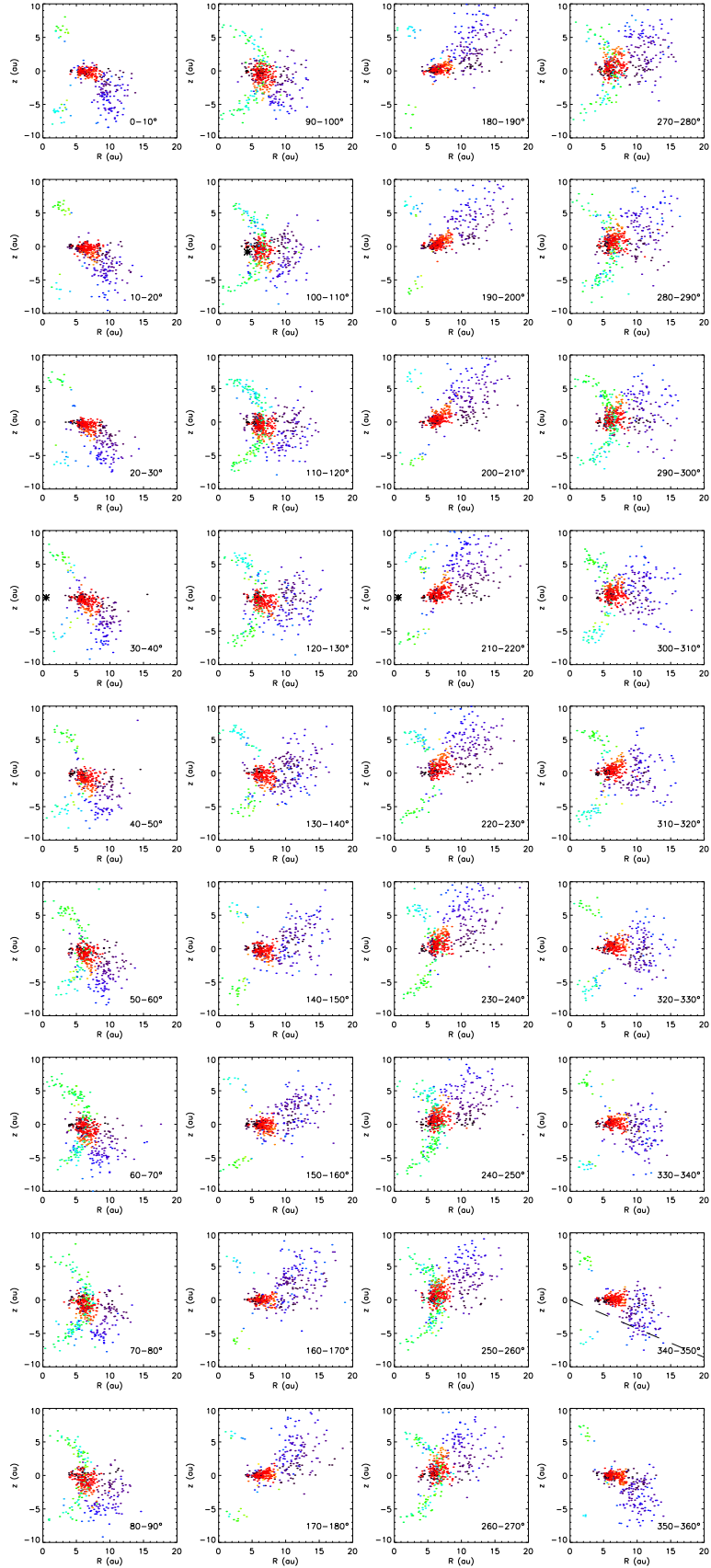}
\caption{
\label{fig:II_rzl}
The azimuthal distribution of material for the simulation using orbit
II. The panels show the radial distance within the plane of the inner
binary and the height above it in segments of $10^\circ$. The
periastron of the inner binary is at $0^\circ$. The stellar positions
are overplotted with the outer star shown reduced by a factor of ten,
and the line of sight shown as a dashed line. The test particle colour
indicates current inner inclination, as in Figure
\ref{fig:II_graphs}.}
\end{figure*}

\begin{figure}
\centering
\includegraphics[width=3.15in]{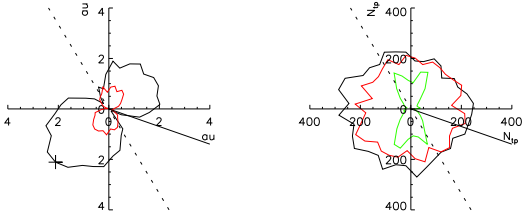}
\caption{
\label{fig:warp}
The azimuthal distribution of particles for the simulation using orbit
II. The left hand panel shows a polar plot of the warp in the prograde
(black) and retrograde (red) discs. The halo is not a disc with
respect to this plane so not plotted. The average height away from the
plane of the inner binary is shown as a function of azimuthal angle,
increasing anticlockwise from the horizontal axis, and the periastron
of the inner binary is at $0^\circ$. The line of sight is shown as a
solid line and the line of intersection of the two orbits as a dashed
line (the rising node is towards the bottom right). A plus marks the
side of the warp that is above the plane. The right hand panel shows a
polar plot of the number of particles in each $10^\circ$ angle
bin. Red and black are as before and the halo is now also plotted in
green.}
\end{figure}

An important feature of the observations of the system is the
extinction and photometric variability towards the B binary,
attributed to disc material along the line of sight by \citet{T99} and
\citet{B05}. \citet{Ak07} find that for the system they investigate
the line of sight can just intercept the top of a warped disc of test
particles. To see what material, if any, occurs along the line of
sight here, we plot the azimuthal structure of the disc, as shown in
Figures \ref{fig:II_rzl} and \ref{fig:warp} for the case of orbit
II. Here, the system has been rotated so that the inner binary lies in
the horizontal plane with its periastron at an azimuthal angle of
$0^\circ$. Each panel then shows the height above the plane as a
function of radial distance within the plane of the inner binary in
segments of $10^\circ$ in azimuthal angle. Particles are colour coded
to inner mutual inclination as for Figure \ref{fig:II_graphs} to
distinguish the different components. The locations of the three stars
are also shown (the position of the outer star is shown reduced by a
factor of ten), as is the line of sight on the relevant plot.

\begin{figure*}
\centering
\includegraphics[width=7in]{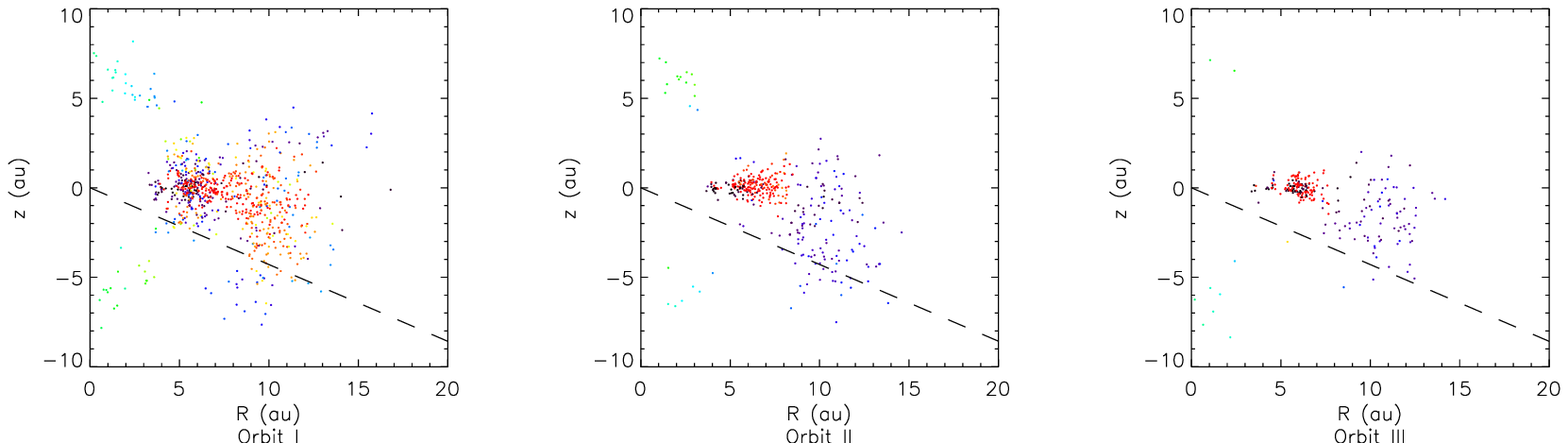}
\caption{
\label{fig:Irz}
The material intercepting the line of sight in all three
simulations. As for Figure \ref{fig:II_rzl} radial distance within the
plane and height above it is shown with plot points colour-coded to
initial inclination. Here however only the segment near the line of
sight has been shown for each simulation.}
\end{figure*}

The warp in the prograde disc is very apparent, with material here
reaching heights of almost 10 au above the plane. The retrograde disc
is slightly less warped, but does not extend out as far. The halo
material shows up very clearly as a ring, instead of a continuous
shell covering all angles. There is a small amount of material
perpendicular to the main ring only at very high $z$, the fourth
population visible in Figure \ref{fig:II_initial}. The warp lies along
the $30^\circ$-$210^\circ$ line, as illustrated in the lefthand panel
in Figure \ref{fig:warp}, but the line of sight at around $160^\circ$
still intercepts a large amount of perturbed coplanar disc material, a
good 5 au above the plane of the binary's orbit. Figure \ref{fig:Irz}
shows similar plots for the other two orbital cases for the azimuthal
segment containing the line of sight. In the high eccentricity case of
orbit III, almost no material is intercepted, while in the low
eccentricity orbit I simulation far more prograde and retrograde
material remains, making either this or orbit II the most likely
orbital configurations if a warped planetesimal disc and associated dust 
is the cause of the observed extinction.

Over the length of the simulation, the warp precesses with a timescale
equal to twice that of the secular period of the stars, following
almost perpendicular to the circulation of the line of intersection of
the two stellar orbits. The extent of the warp in fact decreases as
the mutual inclination between the stars increases (since the orbit is
retrograde the higher the inclination the closer the two planes). The
non-symmetrical distribution seen in Figure \ref{fig:warp} persists,
with one side usually greater in height than the other. The warp is
not a short term feature, and the system will persist in its current
configuration for some time, so if dust follows the planetesimal distribution 
it is very likely the extinction is
indeed caused by the warp in the disc. It should also be noted that
the warped material remains after an additional Myr, and is not
slowly eroded away.

The azimuthal particle distribution shown in the right hand panel of
Figure \ref{fig:warp} shows that the two discs are not skewed in any
particular direction, but that the halo is aligned along the minor
axis of the inner binary's orbit. In fact, it remains aligned at this
angle over the entire simulation. The material in this halo is seen in
projection in the right most panel of Figure \ref{fig:II_faceon} as
the two clumps discussed earlier. As this material follows the
pericentre of the inner orbit, which precesses on the secular time
scale, it will not intercept the line of sight for some time and is
unlikely to be related to the observed variability and extinction.

An important final question raised by \citet{P01} is the lack of a similar
circumbinary disc around the other binary A. As the eccentricity and
mass ratio of this pair are both smaller, a disc should be more likely
here. In fact, the empirical criteria place the stable zone from
around 3 au to 11, 8 and 7 au for orbits I, II and III respectively. A
preliminary simulation taking A as a single star confirms this greater
stability, so the question still remains as to why there is no observed disc
here. It is possible that the inclination of this binary's orbit
places it so that no planetesimals can remain stable, and if so this would
provide limits on the orbit of this stellar pair. Another alternative if there
is indeed a disc of stable planetesimals here is that it is far less dusty and so unobserved.

\section{Conclusion}
\label{sec:conclusion}

Dynamical simulations of a planetesimal population in the debris disc
around the Bab stellar pair in HD 98800 have been run. By studying a
wide range of inclinations, three distinct stable populations have
been identified. These are a prograde disc, a retrograde disc and a
high inclination halo. The radial profiles of each component are
different and distinct. The discs both have large radial gaps, caused
presumably by overlapping resonances from the stellar orbits. The
radial extent of the discs are summarised in Table \ref{tab:rlimits}
but are generally from 3 to 15 au for all three orbits, with gaps at
around 4.5 au and 8 au.

The line of sight can currently pass through slightly warped material
in the prograde disc, and this would account for the observed
extinction and variability if the dust distribution followed the
planetesimals.  However, this alignment with the line of sight
effectively only occurs for the I and II orbits (with eccentricities
of $0.3$ and $0.5$ respectively), which would rule out the higher
eccentricity case ($e = 0.6$) as a possible orbit for the outer star.

The radial profile is complex, and illustrates that, for discs in
multiple systems, dynamical effects purely due to the stars are very
important. Indeed, if most triple stars are within the limit for
resonance overlap to sculpt the stability zone, then this has large
consequences for debris discs and planet formation in such systems.

The simulation results here can be compared to other models of the
system and estimates of the dust distribution. Indeed, as previously
mentioned, comparisons to observations of the dust could place
constraints on the physical processes occuring in the circumbinary
disc. The bulk of the dynamically stable planetesimals are between 5
and 7 au. This matches up well with the prediction of \citet{K00} of a
ring outwards from 5 au. The model of \citet{P01} places the disc from
2 to 5 au, which, apart from the ring around 4.5 au, is unstable
here. \citet{Fu07} suggest an inner ring at 2 au and a thicker puffed
up component at 5.9 au, the only model to predict gaps and similar to
the prograde discs here. The later of these components matches up to
the bulk of the material here, and there is indeed an inner ring seen,
just further out at 4 au.  \citet{P01} estimate the height of the disc
as 1 au, and \citet{Fu07} as 0.75 au. However, here material in the
prograde planetesimal disc is not only warped but very flared,
reaching heights of 5 to 10 au. This is an important dynamical feature
that should be taken into account in models of the
observations. Although roughly agreeing in location, these models do
not have enough detail as yet to further compare to the planetesimals.

The dynamical model of \citet{Ak07} finds stability from 3 to 10 au,
which does not match up that well to the limits here, and in fact
their resulting disc has a very different geometry. Their model uses
different orbital parameters for the stellar system, most notably a
different mutual inclination -- as their aim is not to reproduce the
current configuration but to look at the general morphological
structure of debris discs in highly inclined triple star systems. They
also find that the line of sight can intercept a warp in the disc, but
due to the different geometry it crosses at a different point and at
the maximum warp in the disc, whereas here the warp is only marginally
orientated towards the line of sight. It is in fact material in the
sparser populated outer regions of the prograde disc that intercepts
the line of sight here, from 8 to 15 au, which is a region that in a
coplanar stellar system is predicted to be unstable.

The possible dynamical reason for the high infrared excess already
mentioned in the Introduction is a close pericentre passage of the A
binary stiring up planetesimals and resulting in high collision
rates. Since the binary's period is about 300 years, many such
pericentre passages have occured, and by the end of the simulations
the particles appear very stable with regards to this. For example,
there appears to be no periodic increase in planetesimal
eccentricities as the outer star passes close to the disc. There is,
however, another feature that may cause higher than normal collision
rates. The extent of the flare and warp in the extended disc decreases
as the acute angle between the stellar orbits decreases (and the
mutual inclination increases). As this occurs, planetesimals that have
been spread out in inclination are now packed into less space. This is
likely to result in an increased number of collisions and is
particularly relevant as the current stellar configuration is such
that mutual inclination is large and the disc is near its narrowest
point. However, modelling including collisions is needed to quantify
this effect.

The stable high inclination particles in the halo are curious from a
dynamical perspective. The Kozai mechanism \citep{Ko62} might be
expected to remove such particles very quickly -- however these
particles do not seem subject to this. It is possible that they are in
fact an artifact of the numerical method, but checks with a standard
Bulirsch-Stoer integrator \citep{P89} and the similar observation in
simulations of GG Tau discount this. The high stellar inclinations
involved in the system may be one explanation, but it is worth looking
at the stability should one of the binaries be removed. If the outer
star is not present, a sharp inner disc edge is seen at around 4 au
for all inclinations, with all particles outwards from this remaining
stable. If the inner binary is approximated as a single star instead,
then a reasonably sharp outer edge is seen at around 7 to 10 au, for
the lower inclinations only. The near polar inclinations in this case
are now unstable. The high inclination particles are within this
region which must be shaped by the combined perturbations from both
stellar orbits, so are in a new regime of dynamical behaviour. The
mechanism stabilising these particles, as well as that stabilising the
extended prograde and retrograde discs, is the subject of a future
paper.

Some consideration is needed, however, of how such material would
form. T Tauri multiple systems are common and believed to form
primordially through fragmentation processes, which could result in
non-coplanar discs (\citealt{Ba00}, \citealt{BB00}). As mentioned,
\citet{Mo06} find a small number of multiple T Tauri systems with
misaligned discs in their polarization survey, but suggest that
perhaps these are perturbed disc soon to be realigned. So there is
some evidence that non-coplanar particle distributions are
plausible. It may be possible as well that particles from the disc can
be captured into the polar stability region, and certainly a coplanar
particle distribution is quickly perturbed to fairly high
inclinations.

There is much further work to be done in modelling this system. A
model of the planetesimals including collisions, although expected to
be a minor effect on the dynamics, would indicate if the rate was
currently high and may explain the unually dusty nature of the
system. A model that also included dust and dust collisions, as well
as interactions with any gas in the system, would then fully model the
system and allow detailed comparisons with the observations. Although
not an easy task given the nature of the system it would narrow down
the physics and dynamics at work, important to our understanding of
this stage of the planetary formation process in multiple star
systems.

\section*{Acknowledgements}

We would like to thank Mark Wyatt for bringing this unique system to
our attention, and for helpful discussions with him and Ken Rice. We
would also like to thank the anonymous referee for helpful
comments. PEV acknowledges financial support from the Science and
Technology Facilities Council.



\label{lastpage}

\end{document}